\documentclass[sigconf]{acmart}

\def\BibTeX{{\rm B\kern-.05em{\sc i\kern-.025em b}\kern-.08emT\kern-.1667em\lower.7ex\hbox{E}\kern-.125emX}}
    
\copyrightyear{2019}
\acmYear{2019}
\setcopyright{acmcopyright}
\acmConference[SoICT 2019]{The Tenth International Symposium on Information and Communication Technology}{December 4--6, 2019}{Hanoi - Ha Long Bay, Viet Nam}
\acmBooktitle{The Tenth International Symposium on Information and Communication Technology (SoICT 2019), December 4--6, 2019, Hanoi - Ha Long Bay, Viet Nam}
\acmPrice{15.00}
\acmDOI{10.1145/3368926.3369665}
\acmISBN{978-1-4503-7245-9/19/12}

\begin{document}

%
\title{Systemic Approach for Modeling a Generic Smart Grid}

%

\author{Sofiane Ben Amor}
\affiliation{%
  \institution{Laboratory LI-PaRAD}
  \streetaddress{University of Versailles}
  \city{Versailles}
  \country{France}}
\email{sofiane.ben-amor@uvsq.fr}

\author{Guillaume Guerard}
\affiliation{%
  \institution{Pole Universitaire L\'eonard de Vinci De Vinci Research Center}
  \city{La D\'efense}
  \country{France}}
\email{guillaume.guerard@devinci.fr}

\author{Loup-No\'e Levy}
\affiliation{%
  \institution{Pole Universitaire L\'eonard de Vinci De Vinci Research Center}
  \city{La D\'efense}
  \country{France}}
\email{loup-noe.levy@edu.devinci.fr}

%
\renewcommand{\shortauthors}{Guerard and Levy, et al.}

%
\begin{abstract}
Smart grid technological advances present a recent class of complex interdisciplinary modeling and increasingly difficult simulation problems to solve using traditional computational methods. To simulate a smart grid requires a systemic approach to integrated modeling of power systems, energy markets, demand-side management, and much other resources and assets that are becoming part of the current paradigm of the power grid.

This paper presents a backbone model of a smart grid to test alternative scenarios for the grid. This tool simulates disparate systems to validate assumptions before the human scale model. Thanks to a distributed optimization of subsystems, the production and consumption scheduling is achieved while maintaining flexibility and scalability.
\end{abstract}

%
%


%

%

%
\maketitle

\section{The smart grid's Challenges}
Smart grid started as a fuzzy concept with various definitions which converge on some points \cite{guerard2012survey, shaukat2018survey}. A smart grid is an enhancement of an electrical grid, which attempts to predict and intelligently respond to the behavior and actions of all electric power users connected to it - suppliers, consumers and prosumers (those that do both) - to efficiently deliver reliable, economic, and sustainable electricity services.

The electrical grid has evolved but the growing population and the massive industrial growth of developing countries show the limit of the current grid: increased electricity cost, monolithic infrastructure, losses due to substandard operations at generation, transmission and distribution level. The smart grid is a concept with many features as decentralized management, demand-side management, flexibility, sustainability, resilience, smart services and smart infrastructure.

Many fields of research tend to enhance the grid. Communication technologies, power system management (microgrid, metering, demand response programs) and distributed generation (virtual power plants, integration issues) are keys points of the future grid.

This paper focuses on the power system management -- especially in its simulation. To model a generic smart grid, a systemic approach breaks down the whole system into coordinated sub-problems. The proposed model is a backbone of a smart grid, and one can elaborate some new modules in function of its needs (for example a module to manage small batteries in a smart house). For a overview of existing model, please refer to the following survey published in Energies \cite{uslar2019applying}. And for a non-exhaustive list of applications about smart grid multi-agent models, please refer to the following survey \cite{merabet2014applications}.

This paper is organized as the following: in the next Section, an overview of the model is presented; demand-side management is presented in Section 3 and the demand-response in Section 4; the dynamic structure, built with a topological approach, is described in Section 5; we present in Section 6 the global policies, conducting the overall system behavior; the section 7 and 8 shows the multi-agent model and some results; in the last Section, we conclude about our model.

\section{Overview of the model}
The general process of a smart grid is as follows:
\begin{quote}
The bidirectional power and communication flow will help to buy and sell electrical energy from utility, as a server and client \cite{fernandes2010identification}. The server receives a request for energy supply, and on approval, client will receive energy.  The consumers can directly supervise energy by monitoring their energy usage profile \cite{sa2009electricity}. 
\end{quote}

Taking into account its structure, its goals and its needs, a smart grid is breaking down into three subsystems with their own behaviors: the T\&D network, the microgrid and the local layer. These three subsystems form the backbone of any energy grid.

\begin{description}
\item[T\&D:] transmission and distribution grids conduct electricity to consumers from a fully connected grid to a linear path, at the same time as balancing supply with demand.Their missions are to ensure an equitable and non-discriminatory access to the networks. Algorithms at this level have to limit the effects of congestion due to the widespread use of a few lines while limiting the losses of energy during the routing.

\item[Microgrid:] the microgrids ensure for a district, a rural area or a group of renewable energies to link with the energy market and the distribution network. Its role is to establish a consensus between consumption and production, like an aggregator.

\item[Local:] the last subsystem represents the end nodes of the network. It models prosumers, i.e. a group of consuming devices, local renewable plants or/and electric vehicle, requesting and providing a limited amount of energy.
\end{description}

\begin{figure}[!ht]
\centering
\includegraphics [width=0.45\textwidth]{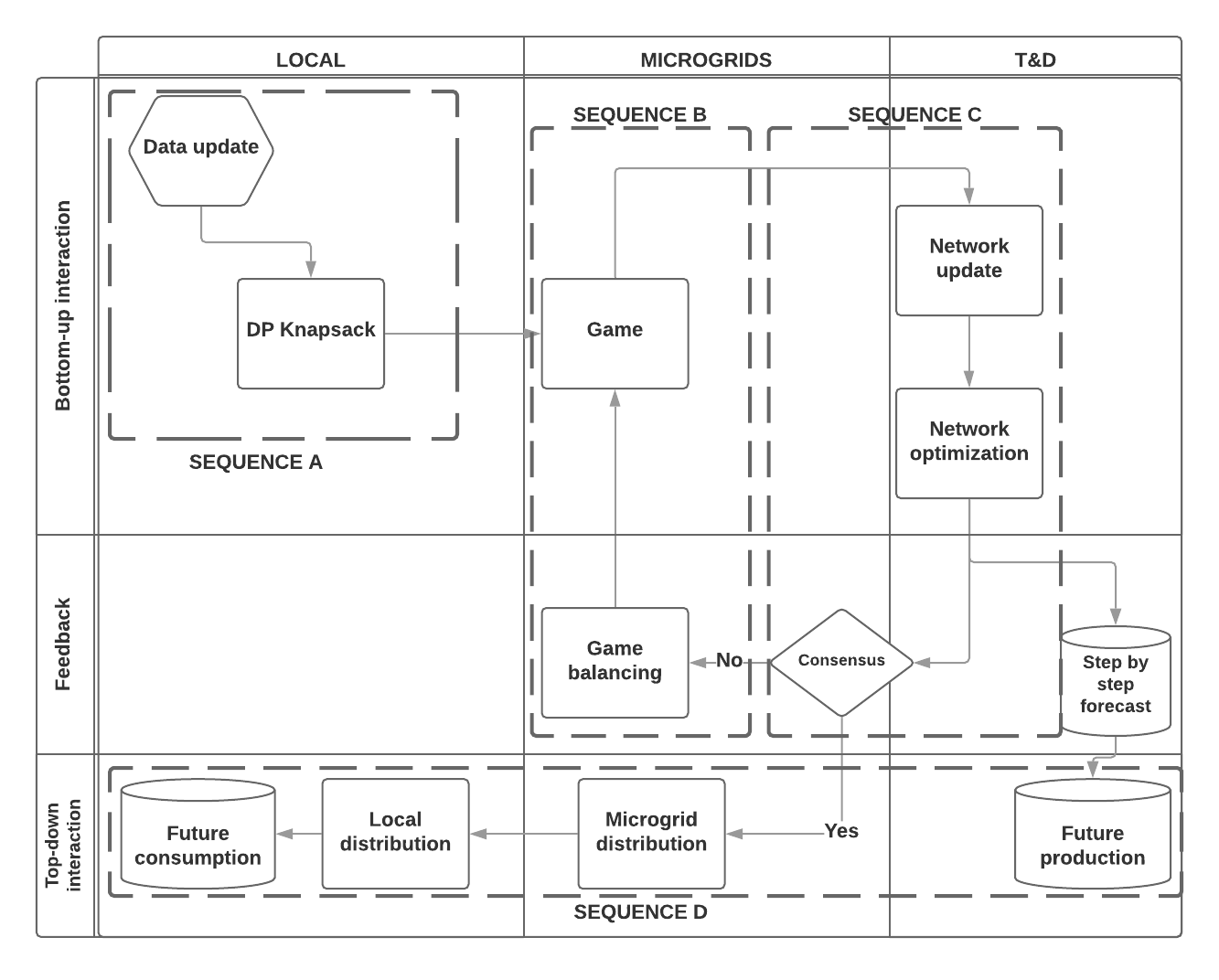}
\caption{Sequential Scheme.}
\label{fig4}
\end{figure}

The proposed model is based on discrete time, i.e. each agent of the model is based on the same timer. An iteration happened with a constant time interval. The process follows four sequences (figure \ref{fig4}) as follows:

\begin{description}
\item[A new iteration begins.] 
\item[Sequence A:] each prosumer develops strategies of consumption thanks to a dynamic knapsack problem.

\item[Sequence B:] each microgrid establishes a game taking into account prosumer's strategies and response's strategies. A strategy based on pareto equilibrium is chosen for each prosumer.

\item[Sequence C:] energy flows are optimized across the grid. Each microgrid identifies the amount of energy received and must adjust their bid following a feedback to Sequence B.

\item[Sequence D:] when a consensus is established after some sequences B--C loops, each prosumer and producer establish their new forecast and consumption's strategies.

\item[End of the iteration.] 
\end{description}

Any new steps or sequences can be grafted to the backbone by a direct insertion into the process, provided that data's I/O do not disrupt the whole process. All sequences are explained in detail in the following sections.

\section{Local's Management: \textit{sequence A}}\label{demand}
A prosumer is assimilated as a group of consuming devices, renewable energies and electric vehicles requesting or providing a measurable amount of energy. Each device can be managed individually.

\subsection{Process}
To manage home automation, a variable is added to each device, reflecting its necessity to consume. This variable, called \textit{priority}, is a key to demand-response management. A priority equals to zero means the device consumes energy. The higher the priority, the less the device needs energy.

Some of the devices may respond to a direct control, i.e. some sensors that indicates how to consume. Those devices are not includes in any demand-response program. Indeed, those devices, as brightness control, security system or blinds, adapt their behaviors to some direct stimuli and require an immediate source of energy. Their priority value is always equal to zero, but they may not consume any energy.

Other devices, through sensors and internal parameters, may change their priority value over time. A room with a thermostat computes the evolution of its temperature. In function of human behavior and its comfort, a smart management adjusts the consumption of heating/cooling system.

Over here is explained the process of the sequence A:

\begin{description}
\item[Step 1] Data Update.%

\item[Step 2] Knapsack: a knapsack problem is resolved to build a set of consumption's schemes.

\item[Step 3] Dynamic programming: the knapsack subsolutions are stored to compute various schemes of consumption.
\end{description}

\subsection{Knapsack problem}
The knapsack problem is a problem in combinatorial optimization defined by Tobias Dantzig in Numbers: The language of Science in 1930. The most widespread problem being solved is the 0-1 knapsack problem \cite{martello2000new}, which restricts the number $x_i$ of copies of each item to zero or one. Given a set of $n$ items numbered from $1$ up to $n$, each with a weight $w_i$ and a value $v_i$, along with a maximum weight capacity $W$, the knapsack problem is:

$$
\left\{
\begin{array}{l}
\mbox{ maximize } \;\sum_{i=1}^n v_i x_i,\\
\mbox{ subject to } \sum_{i=1}^n w_i x_i \leq W and \; x_i \in \{0,1\}
\end{array}
\right.
$$ 

Informally, the problem is to maximize the sum of the values of the items in the knapsack so that the sum of the weights is less than or equal to the knapsack's capacity.
    
The problem is solved by dynamic programming. Assume $w_1$, $w_2$, $\ldots$ , $w_n$, $W$ are strictly positive integers. Define $m[i,w]$ to be the maximum value that is reached with weight less than or equal to $w$ using items up to $i$. We define $m[i,w]$ recursively as follows:
   $$
   \left\{
\begin{array}{l}
m[0,\,w]=0,\\
 m[i,\,w]=m[i-1,\,w] \; if \; w_i > w\,\!, \\
 m[i,\,w]=\max(m[i-1,\,w],\,m[i-1,w-w_i]+v_i) \\
 \qquad\qquad\qquad\qquad\qquad\qquad\qquad\qquad\;if\; w_i \le w
\end{array}
\right.
 $$
The solution is found by calculating $m[n,W]$. To achieve this efficiently, a table is used to store previous computations. All optimal solutions are stored in this table. 

Our problem is well-suited with the 0-1 knapsack problem. Let $n$ be the devices in a smart house, each device consume $w_i$ energy at the current time. To know all the schemes of consumption of the smart house for the current time, the problem is unbound, i.e. $W$ as an infinite value -- the problem stop when all the items are used in the knapsack, i.e. each $x_i = 1$ for $i=1\ldots n$. 

A function computes for each device a value, according to its priority and its consumption. Values of device $i$, noted $v_i$ in each house follow this method: let $w_{max}$ be the greatest consumption in the house, $p_{max}$ be the greatest priority in the house; for each device $i$ in the house, $v_i = (w_{max} *  p_{max}) - (w_i *  p_i) + w_i$.

Once the table for the 0-1 knapsack problem is complete, we have to compute another dynamic problem to identify which devices are used for any $w$ in $m[n,w]$. Starting at the position $m[i,w]$ where $i=n$:
\begin{itemize}
\item if the value of $m[i,w]$ is different to $m[i-1,w]$ then the device $i$ is used in the solution; set $m[i,w]$ to $m[i-1,w-w_i]$
\item if the value of $m[i,w]$ is equal to $m[i-1,w]$ then the device $i$ isn't used in the solution; set $m[i,w]$ to $m[i-1,w]$
\item restart the process until $w=0$
\end{itemize} 
In this manner, when a strategy is selected during sequence B, the prosumer recognizes which devices consume.

\section{Microgrid's management: \textit{sequence B}}\label{microgrid}

Once each prosumer has developed its consumption's strategies, they communicate their result to the microgrid controller. At that point, the process occurs in two steps as follows: 

\begin{description}
\item[Step 1] Auction: the microgrid experiences the set of consumption's strategy from each smart house. A game regroups each strategy of consumption with distribution strategies. On that occasion, an equilibrium of Pareto is obtained.The sum of each chosen house's strategy of consumption represents the total consumption for the microgrid or the energy that must be provided. 

\item[Step 2] Feedback: once the T\&D network has computed the routing, the smart grid is able to find which microgrids ask too much energy and those which can inquire more energy. Feedbacks punish or reward each local level to perform an improved auction.The purpose of the feedback system is to balance supply and demand without operating a deterministic mathematical system, allowing a flexible network. The feedback system is described in section \ref{global}.

\item[Step 3] Final allocation: when the supply and demand are balanced, each smart house utilizes the dynamic program of knapsack problem to find the optimal distribution solution according to the chosen strategy.
\end{description}

\subsection{Game theory}
The negotiation of technical arrangements must take into account that each member of a microgrid, both consumers and producers, is motivated to maximize its own profit. Game theoretic reasoning pervades economic theory and is employed widely in other social and behavioral sciences. Briefly, this theory is a decision-maker selects the worthiest action according to its preferences, among all the actions present to it. For a thorough introduction to game theory, we refer to the book of M.J. Osborne \cite{osborne2004introduction}.

A strategic game is a model of interacting decision-makers, also called players. It is defined as follows: a set of players; for each player, a set of strategies; for each player, preferences over the set of strategy profiles (a value for each strategy). 

Time is absent from the model. The idea is each player decides on its action once and for all, and the players select their actions simultaneously in the sense that no player is informed, when one decides on its action, of the action chosen by any other player.

A two-player strategic game is typically represented as the Table \ref{SchemeGame} where each row follows a consumption's strategy for a smart house and each column follow a response strategy from the T\&D network.

\begin{table}[!ht]
\begin{center}
\caption{Scheme of a game between a smart house and producers}
\begin{tabular}{|r|c|c|c|}
\hline
House/T\&D & Response & \ldots &  strategies \\
\hline
DSM & \ldots /\ldots & \ldots /\ldots & \ldots /\ldots \\
\vdots & \ldots /\ldots & \ldots /\ldots & \ldots /\ldots \\
strategies & \ldots /\ldots & \ldots /\ldots & \ldots /\ldots \\
\hline
\end{tabular}
\label{SchemeGame}
\end{center}
\end{table}

The junction of a raw and a column gives two payoffs, one for each player. In the proposed model, the most appropriate choice (junction) is a Pareto optimum, i.e. an economic decision. Pareto optimality represents a state of allocation of resources from which it is impossible to relocate so as to make any one individual or preference criterion better off without making at least one individual or preference criterion worse off \cite{rabin1993incorporating}.

We have to calculate the value for prosumer and distribution for each strategy of each house. The value $l$ for the prosumer is equal to $l=\sum_{i=1\ldots n}\frac{u_i * w_i}{p_i}$. For the distribution, the value of a strategy depends on the average of utilities for an unit of energy in the smart house, $\gamma = \sum_{i=1\ldots n}\frac{u_i * w_i}{w_i}$, the value $r$ is equal to $r=\sum_{i=1\ldots n}(\frac{u_i}{p_i}-\gamma)w_i$. The key idea behind those two functions is to strike an appropriate balance between the priority of consumption and the amount of energy.

\subsection{Rewards and punishments}
After a game, the T\&D network looks for the best routing of energy. Some tests identify the bottlenecks or the mismanagement of energy. Then, those tests submit a feedback to each microgrid about their consumption.

It supports only three messages: $(a)$ one may consume less energy, $(b)$ one fits with the routing, $(c)$ one can ask for more energy.

If a microgrid receives a message $(a)$, then for each smart house, all the strategies that consumes more or equals to the current one have their value increased. The opposite applies to a message $(c)$. If a microgrid receives a message $(b)$, then nothing happens.

To increase or decrease the value, a coefficient $\epsilon$ is applied to them. To decrease the value, it is multiplied to $1-\epsilon$; to increase the value, it is multiplied to $1+\epsilon$ with $0<\epsilon<<1$.

\section{A Dynamic Structure: \textit{sequence C}}\label{dynamic}
The electricity flow which is sent to any microgrid not merely depends on the production but also to the distribution. Since each microgrid bid for an amount of energy, the transmission and distribution network (T\&D) must check if production and consumption match and if energy can be routed from producers to prosumers.

Traditionally, this is resolved with the help of automatic voltage regulators and using supervisory control and data acquisition systems \cite{chowdhury2009microgrids}. Using such systems helps to configure the network and to send control signals to actors to increase or decrease production or consumption.

The algorithm at T\&D level has to limit congestion, while limiting the energy losses by Joule heating. If the process recognizes any misconduct or consumption and production unmatched, the algorithm analyzes bottlenecks to adjust the bids thanks to a feedback. A maximum flow problem is consistent with these criteria.

\subsection{Process}

T\&D level receives energy requested by each microgrid. Electricity has to be routed from producers to these microgrids while preventing congestion. The process is:

\begin{description}
\item[Step 1]Residual Network: production or consumption may change between each iteration or feedback. The graph updates its value.
\item[Step 2]Routing: the energy routing is calculated by a max flow problem (Busacker \& Gowen Algorithm \cite{busacker1960procedure}). Once routing is achieved, the data are recorded to predict possible variations of the subsequent iterations. 
\item[Step 3]Feedbacks: it is likely that producers do not provide all available energy, or microgrids do not receive all energy needed. An algorithm aims to analyze gaps or bottlenecks in the T\&D network to send feedbacks to microgrids.
\end{description}

\subsection{How to construct a complex network}

\begin{figure}[!ht]
\centering
\includegraphics [width=0.25\textwidth]{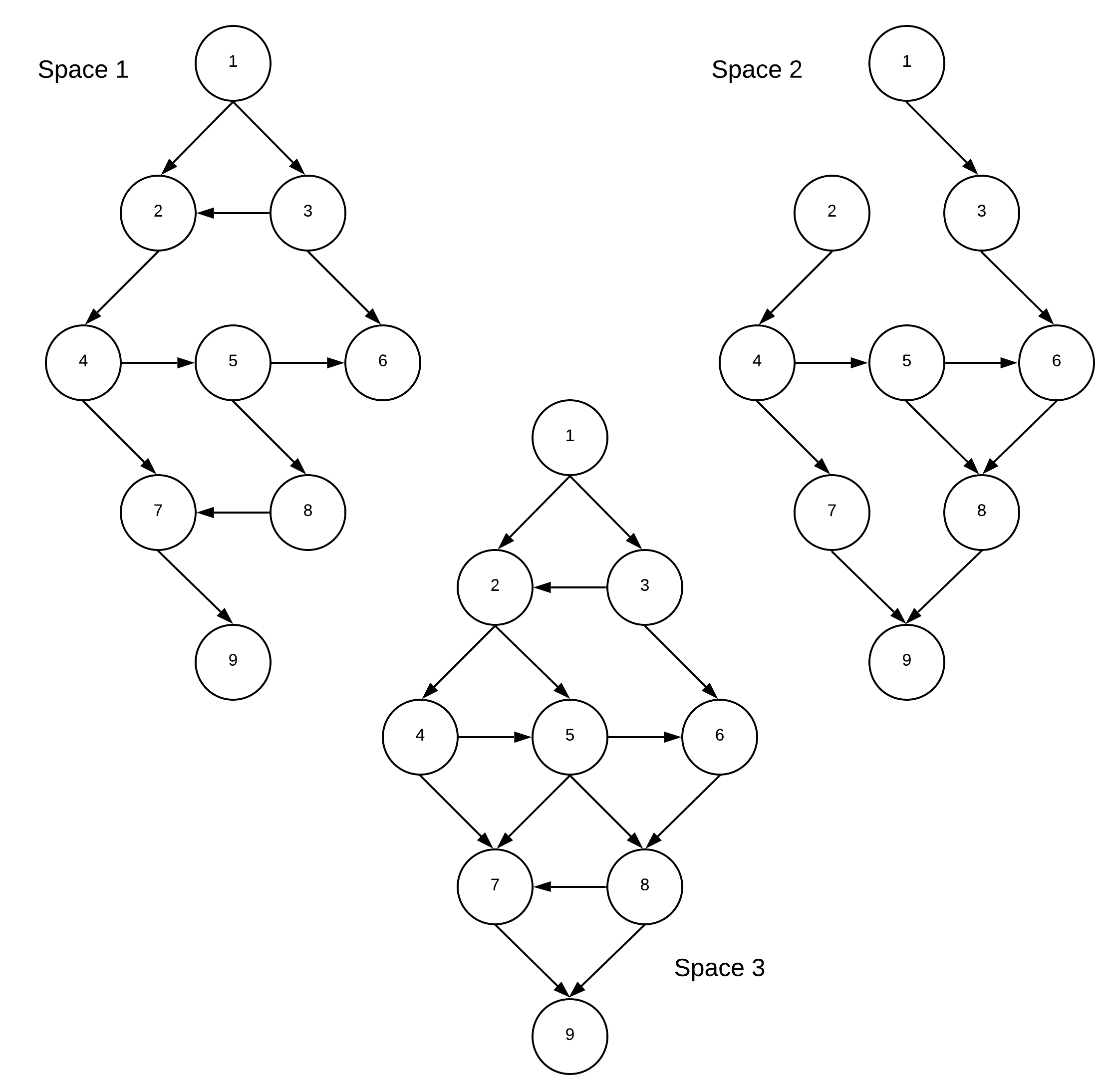}
\caption{Family of topologic spaces.}
\label{Pretopologic}
\end{figure}

\begin{figure}[!ht]
\centering
\includegraphics [width=0.20\textwidth]{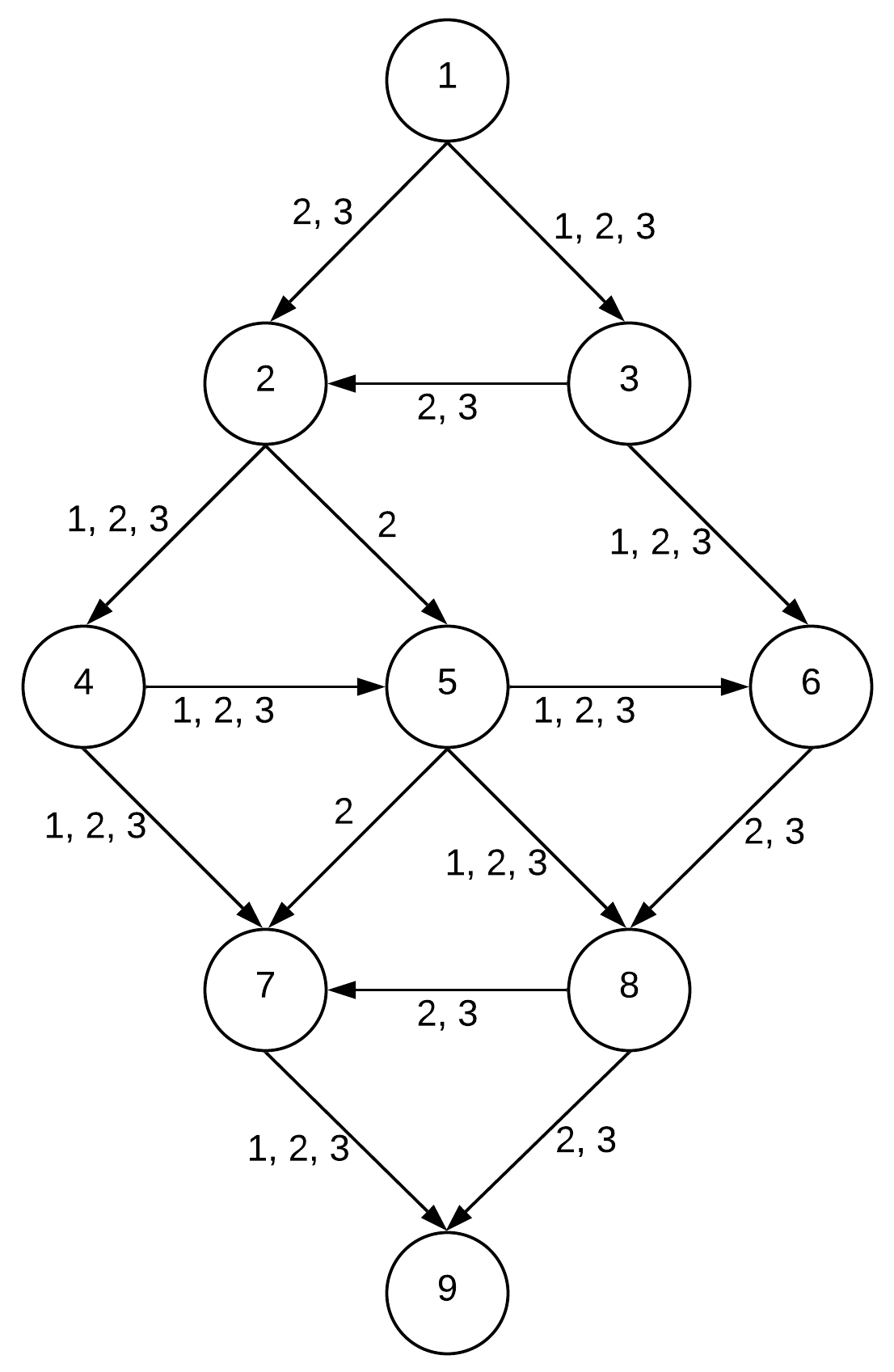}
\caption{Network according to topologic spaces.}
\label{Graphe-finale}
\end{figure}

The amount of consumption for each microgrid and the amount of production in the grid change over time. In our discrete model, both consumption and production change from an iteration to the following. To take into consideration those change, and also the variability of the line's capacity, the network is seen as a family of topologic spaces. The overall adhesion function, constituting an alternative space, is defined as an aggregation of several topologic spaces. The adhesion function doesn't need the know the whole network since the function only needs to know its neighbors. In this manner, a modification happens in only one space, but have an impact on the model. This method is affiliated to the pretopology method \cite{stadler2002recombination}.

For example, let be three topologic spaces $a_{1}$, $a_{2}$ and $a_{3}$. Each edge possesses three levels of flow corresponding to under-load, standard load and over-load.
The figure \ref{Pretopologic} presents the topologic family.
Under-load (1) is possible on an edge if it exists in the following logical space $a_{1}\cap a_{2}\cap a_{3}$, same for over-load (3) in the logical space $(a_{1}\cup a_{2})\cap a_{3}$. By default, any edge of Space 3 carries an average standard load (2). The figure \ref{Graphe-finale} presents the final graph.

\subsection{Routing problem}
The previous subsection presents how to build a graph according to several topologic spaces. About the routing problem or min cost flow problem, nodal rule or Kirchhoff’s circuit specifies that at any node in a circuit, the sum of currents flowing into that node is equal to the sum of the currents flowing out of that node. An electrical circuit is similar to a graph in which a junction is a node, and physical connection corresponds to an edge.

Each edge $(u,v)$ is characterized by:
\begin{itemize}
\item $d(u,v)$
the maximum capacity;

\item $l(u,v)$
the minimum capacity;

\item $c(u,v)$
the unit cost of the flow in the edge. The cost may vary in function of the total flow. An edge is duplicate with various costs related to capacity. The cost function models the Joule losses.
\end{itemize}
  
From the example of the previous subsection, because the cost function is strictly increasing in function of flow amount, this method does not disturb the algorithm of maximum flow at minimum cost. For example, the cost of one unit of flow on an edge tagged 1 (under-load) is equal to 1 -- arbitrary value; the cost of one unit of flow on an edge tagged 2 (standard load) is equal to 2; the cost of one unit of flow on an edge tagged 3 (overload) is equal to 4.

A path for a producer to a microgrid with an available capacity is called an augmenting path. At each iteration, the edge $(u,v)$ is valued at $c(u,v)$ if the edge $(u,v)$ is unsaturated. The edge $(v,u)$ is valued at $-c(u,v)$ if the edge $(u,v)$ is not empty; let $f(u,v)$ be the flow passing through the edge.

To have a unique source and a unique sink, new nodes are connected. All producers, including microgrid which exchange energy, are linked to a virtual node named $source$ : $s$. The capacity of the edge is equal to the amount of energy produced. Respectively with the consumers where the capacity of the edge is equal to the energy bid, the virtual node is called $sink$ : $t$. The flow is routed from the source $s$ to the sink $t$.

To resolve this routing problem, the Busacker \& Gowen algorithm is used. The idea behind the algorithm is: as long as there is a path from the source   to the sink, with an available capacity on all edges in the path, we send flow along one of these paths, filling in priority the path with the minimal cost. Then another path is determined, and so on. 

For a mathematical use, the problem is as follows (based on network simplex):

\begin{tabular}{ll}
 minimize & $\sum_{(u,v) \in E} c(u,v) \cdot f(u,v)$ \\
 subject to &  \\
& $l(u,v) \le  f(u,v) \le d(u,v)$ \\
& $f(u,v) = - f(v,u)$ \\
& $\sum_{w \in V} f(u,w) = 0$ for all $u \neq s, t$ \\
& $\sum_{(s,u) \in E} f(s, u) = \sum_{(v,t) \in E} f(v, t)$ \\
\end{tabular}

\noindent where $G=(V,E)$ is a directed graph, with source $s \in V$ and sink $t \in V$.

\subsection{Updating the network}
Since the Busacker \& Gowen algorithm is solved by dynamic programming, a current solution is deducted from previous iterations or solutions. An adjustment algorithm balances the graph with the current values for all producers and microgrids.

Let $d^{*}(u,v)$ be the difference between the old and the capacity of the edge $(u,v)$, this edge is listed. Only vertexes with a lower flow capacity at the previous iteration are taken into account. This is also executed in the case of congestion on a line.

Once the edges are listed, a new graph $G'$ is created where each capacity is given by: $d(u,v)=f(u,v)$ the previous flow passing through the edge $(u,v)$; or $d(u,v)=d^{*}(u,v)$ if $(u,v)$ is in the list. All costs are inverted: $c(u,v)\leftarrow \frac{1}{c(u,v)}$.   

The Busacker \& Gowen algorithm is executed on $G'$ until all listed edges are saturated. As long as there is an edge $(u,v)$ unsaturated in the list, the algorithm continues. The result is substrate to the graph $G$. 

Finally, the Busacker \& Gowen algorithm is applied on $G$ updated. The result gives the routing.

\section{Global policies: \textit{feedback and sequence D}}\label{global}
We have seen in the previous section how smart houses develop consumption's strategies, how a microgrid bid for some energy and how the T\&D network routes energies from producers to microgrids. But at this point, a feedback system is needed to ensure a consensus between all the agents of the system and to validate the current iteration.

As a complex system, a smart grid needs global policies or goals to ensure its health. Policies are executed to avoid inequality, local disturbance, undesirable local behavior and to adjust its forward strategies' choice.

\subsection{Feedbacks}
We note that sources or sinks may be unsaturated after the routing process. This is a mismanagement of resources at the consumer or mismanagement of the energy produced. An algorithm reveals the overused or underused nodes to perform feedback. Two tests are performed: 

\begin{itemize}
\item The min cost flow problem on the graph without capacity constraints for microgrids. The result will provide the maximum unconstrained consumption. 

\item The min cost flow problem on the graph without capacity constraints for producers. The result will reflect consumer demand to predict forthcoming production.

\end{itemize}

The problem of maximal flow contains multiple valid patterns. Feedback reorganizes the distribution of resources among various microgrids. They will adjust their demand until another feedback. 

The gap between the constrained solution and the two tests determined how to perform the feedback. The values obtained by the graphs are used in the feedback to punish or reward smart house's strategies \cite{camerer2003behavioral}. Both tests should be performed at each feedback to take into account the results of the new bids.

\subsection{Short term forecast}
To prevent brownouts and blackouts, it is relevant to identify the users' behavior. Forecasts provide a significant impact on the running of the smart grid. Indeed, a smart grid aims to smooth consumption curve while ensuring energy supply. 

When searching for the shortfall in production or consumption during the sequence C, the ideal distribution of production without consumption constraint and the ideal distribution of consumption without production are calculated (see previous section). They deliver valuable data to follow the evolution of consumption over time. The forecast is calculated at the end of the sequence D.

The forecast is calculated by a weighted average of the bids conducted. Let $z_{i}$ be the bid made at the feedback $i-1$, the forecast is calculated as follows: $Z=2\frac{\sum_{i=1}^{n}i*z_{i}}{n(n-1)}$, the latest bids produce a more significant impact on the forecast. This forecast is applied to each microgrid.

\begin{figure}[!ht]
\centering
\includegraphics [width=0.35\textwidth]{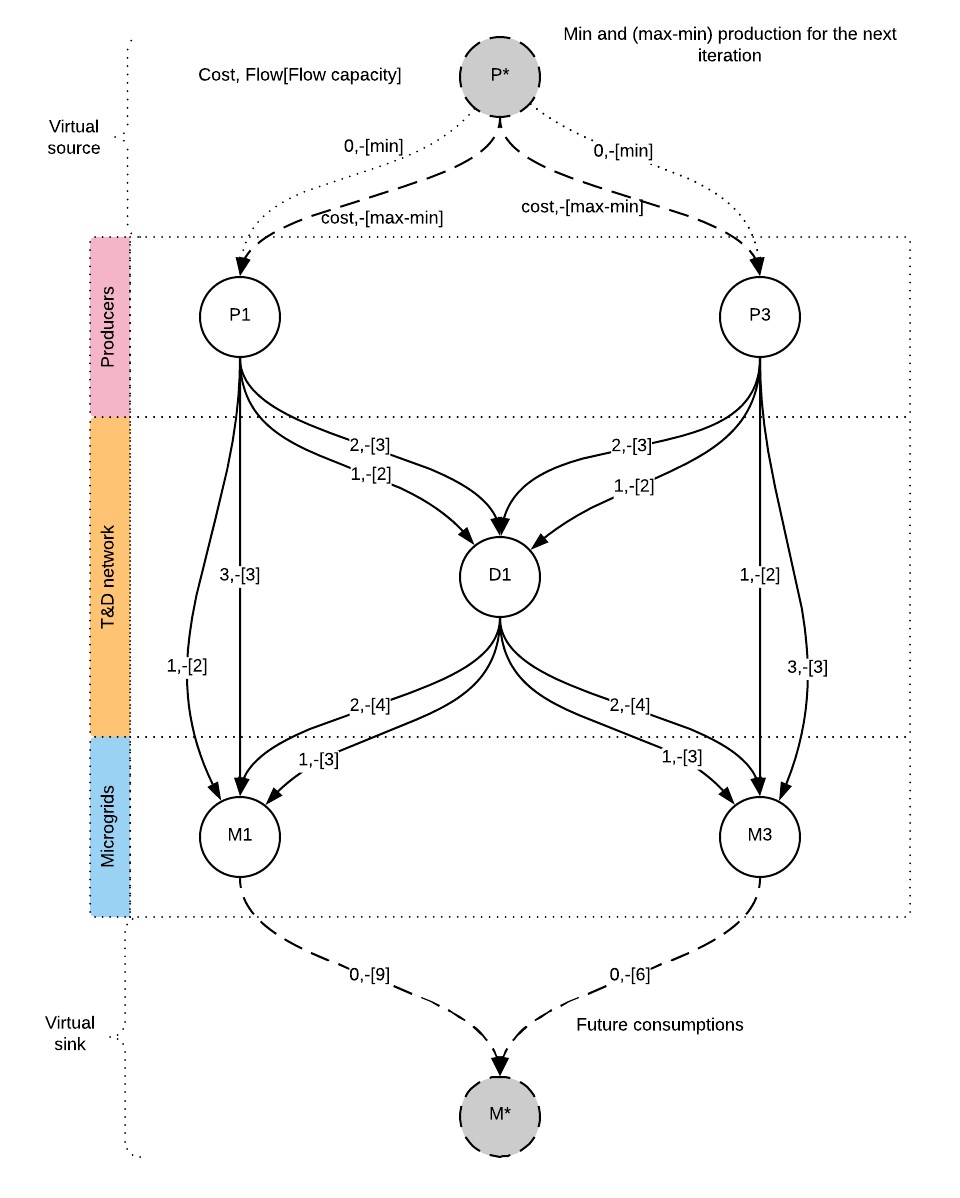}
\caption{Forecast for microgrid's consumption and plants' production.}
\label{future}
\end{figure}

For the producers, we know the future consumption value and also known the cost of a variation of production for each plant and the available variation of the amount of energy produced. A routing problem is resolved to determine how producers must amend their production as described in figure \ref{future}.

\section{Multi-agent model and consumer's behaviors}
The multi-agent simulations are made with GAMA. The GAMA platform includes various aspects as a grid environment, GIS integration, multi-level modeling and equation-based models which sue with a smart grid simulation.

\begin{figure}[!ht]
\centering
\includegraphics [width=0.45\textwidth]{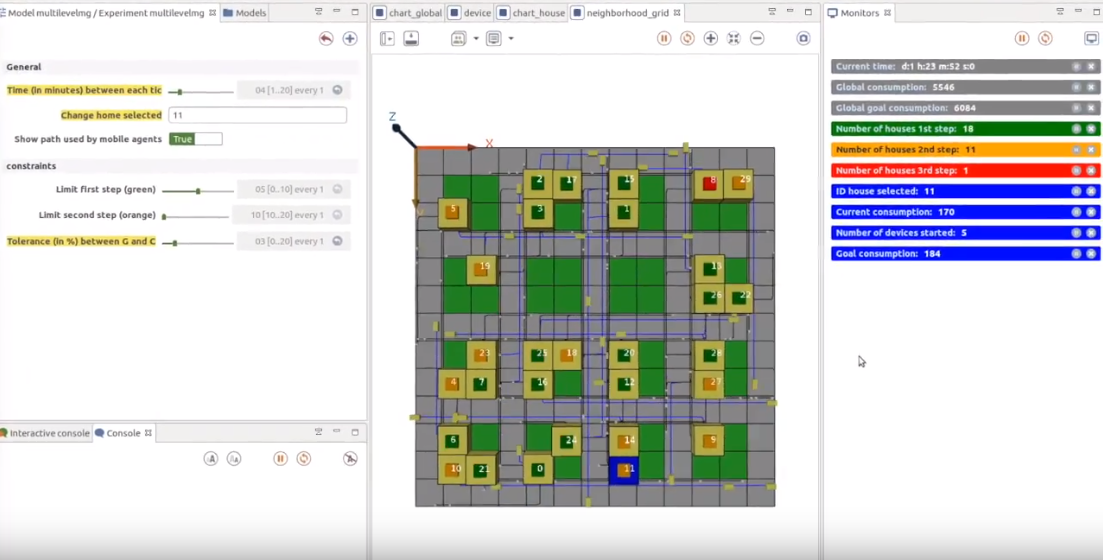}
\caption{Simulation of a microgrid.}
\label{gama1}
\end{figure}

The GAMA model can simulate various microgrids. Each microgrid is defined by its number of smart houses and its goal (i.e. the amount of energy from the producer after a game). Each smart house has a random number of devices (with various behaviors). The simulation shows in green, orange or red if the smart house follows a good trend or involves more energy than expected. The figure \ref{gama1} presents a small microgrid simulated with GAMA.

\begin{figure}[!ht]
\centering
\includegraphics [width=0.45\textwidth]{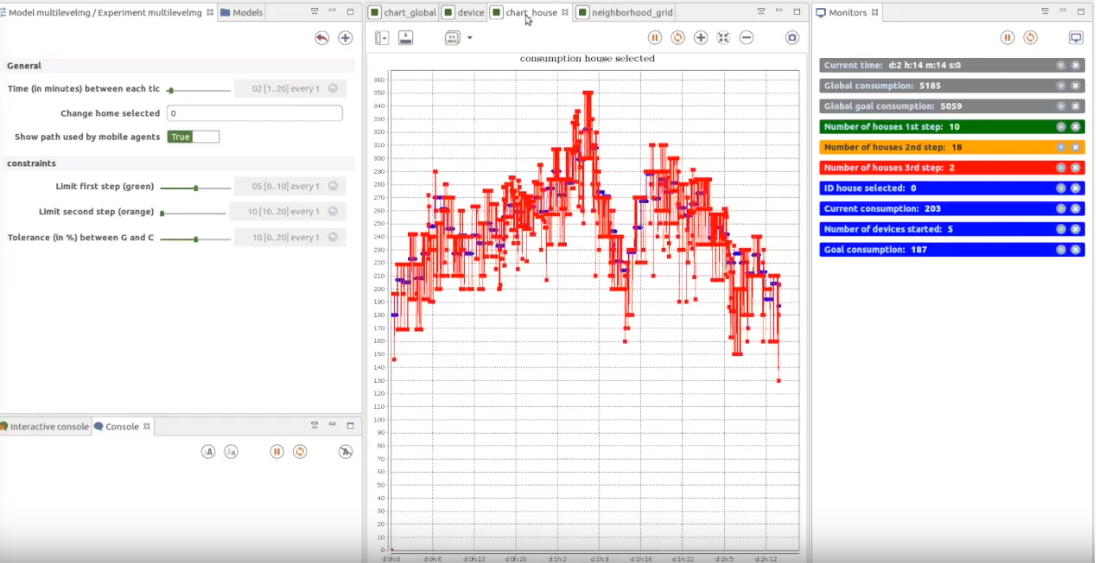}
\caption{Consumption of a single smart house.}
\label{gama2}
\end{figure}

\begin{figure}[!ht]
\centering
\includegraphics [width=0.45\textwidth]{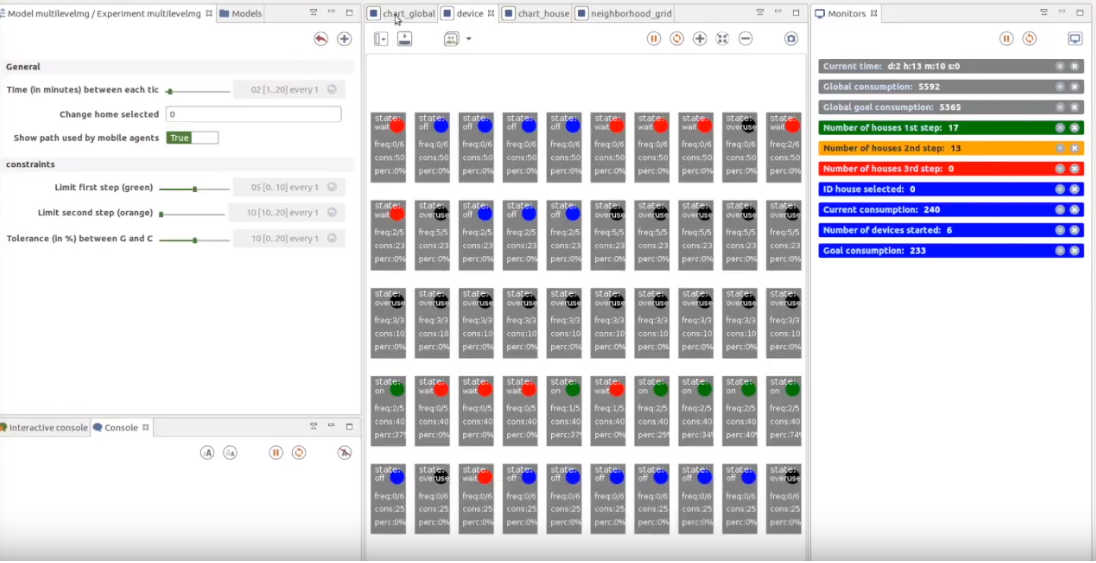}
\caption{State of each devices of a smart house.}
\label{gama3}
\end{figure}

The figures \ref{gama2} and \ref{gama3} respectively show the consumption and the devices of a smart house. The consumption curve presents the prediction in blue and the final amount of consumed energy in red. The curve can be simulated for an unique house or for multiple ones, from the same microgrid or not.

For the devices, the color is in function of its state. Since there are multiple behaviors, colors present if the device is in use (in green), if the device will no longer be utilized during the simulated day (in black), as a decreasing priority or an increasing priority (resp. in blue and in red).

\section{Results}
\label{sec_results}

\subsection{Step by step example}
The iteration starts with the sequence A. The Table \ref{Game0} presents a set of devices for three smart houses with in order: the device ID, energy/priority of consumption/local knapsack value. The prediction shows that the first house needs 10 amounts of energy (arbitrary value), the second one needs 7 and the last one need 12. The result of the knapsack problem includes the devices in bold.

\begin{table}[h!]
\caption{Set of devices for each smart houses}
\begin{tabular}{cccc}
Smart house & 1       & 2      & 3     \\
Device 1    & \textbf{1/0/81}  & \textbf{1/0/16} & \textbf{1/0/-} \\
Device 2    & \textbf{1/1/80}  & \textbf{1/0/16} & \textbf{1/0/-}   \\
Device 3    & \textbf{3/0/83}  & \textbf{2/1/15} & \textbf{10/0/-} \\
Device 4    & \textbf{5/2/75}  & \textbf{3/0/18} &        \\
Device 5    & 20/4/20 & 4/3/7  &        \\
Device 6    &         & 5/3/5  &             
\end{tabular}
\label{Game0}
\end{table}

Then, the sequence B builds some strategies of consumption. In this example, a first strategy takes the devices with a priority equals to 0, then up to 1 et caetera. The Table \ref{Game1} shows the set of strategies for each smart house.

\begin{table}[h!]
\caption{Set of strategies for each smart house}
\begin{tabular}{cccc}
Smart house & 1                    & 2                     & 3              \\
Strategy 1  & 4/\{1,3\}/164        & 5/\{1,2,4\}/50        & \textbf{12/\{1,2,3\}/-} \\
Strategy 2  & 5/\{1,2,3\}/244      & \textbf{7/\{1,2,3,4\}/65}      &                \\
Strategy 3  & \textbf{10/\{1,2,3,4\}/319}   & 11/\{1,2,3,4,5\}/72   &                \\
Strategy 4  & 30/\{1,2,3,4,5\}/339 & 16/\{1,2,3,4,5,6\}/77 &                        
\end{tabular}
\label{Game1}
\end{table}

Now, we have to calculate the value for prosumer and distribution for each strategy of each house to build a game.

The Table \ref{Game2} presents the value $l/r$ for each strategy of each house. A Pareto equilibrium will select the strategies in bold for every smart house.

\begin{table}[h!]
\caption{Game between smart house and producers}
\begin{tabular}{cccc}
Smart house & 1        & 2       & 3   \\
Strategy 1  & 330/194  & 77/27   & \textbf{-/-} \\
Strategy 2  & 410/240  & \textbf{107/37}  &     \\
Strategy 3  & \textbf{620/257}  & 125/-36 &     \\
Strategy 4  & 720/-322 &         &    
\end{tabular}
\label{Game2}
\end{table}

\begin{figure}[!ht]
\centering
\includegraphics [width=0.33\textwidth]{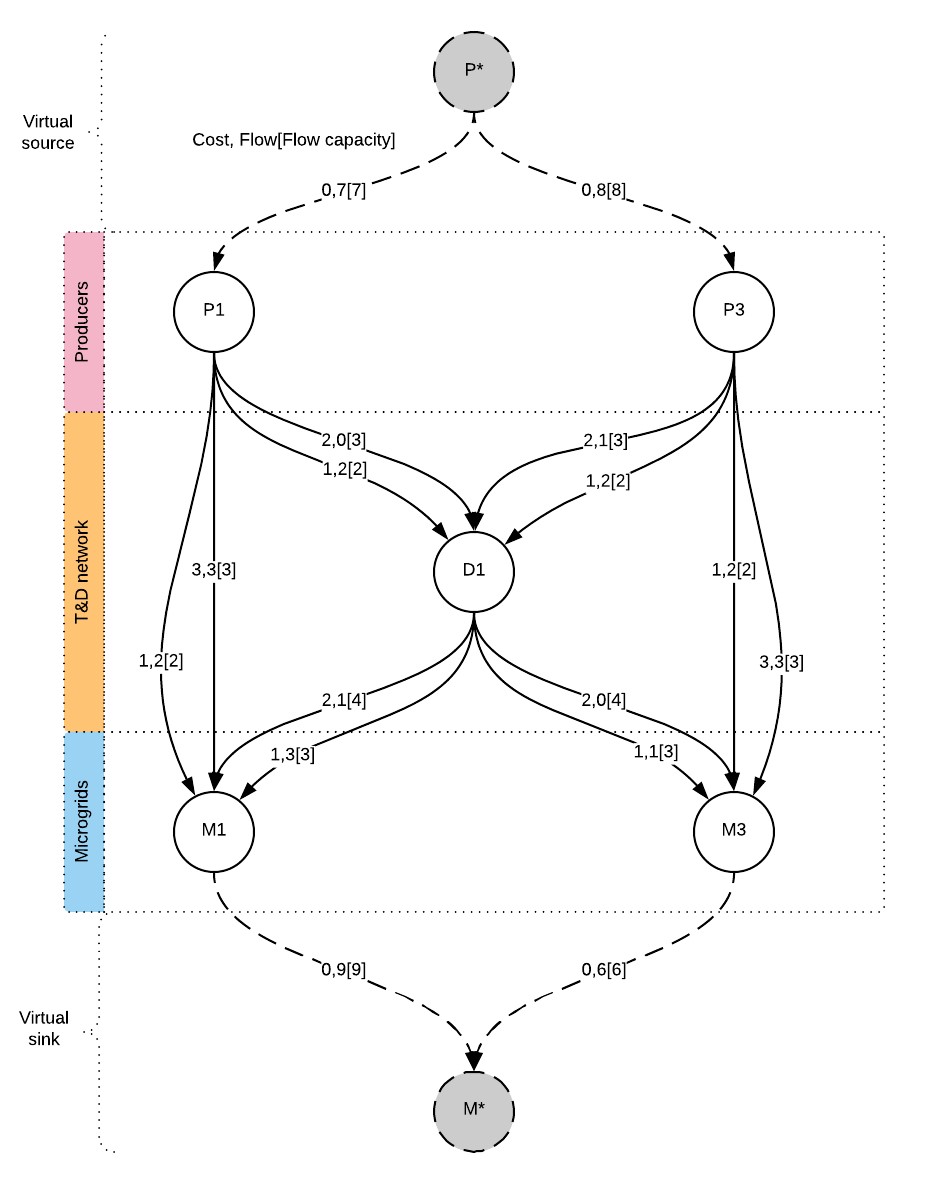}
\caption{Previous routing.}
\label{rourou1}
\end{figure}

\begin{figure}[!ht]
\centering
\includegraphics [width=0.33\textwidth]{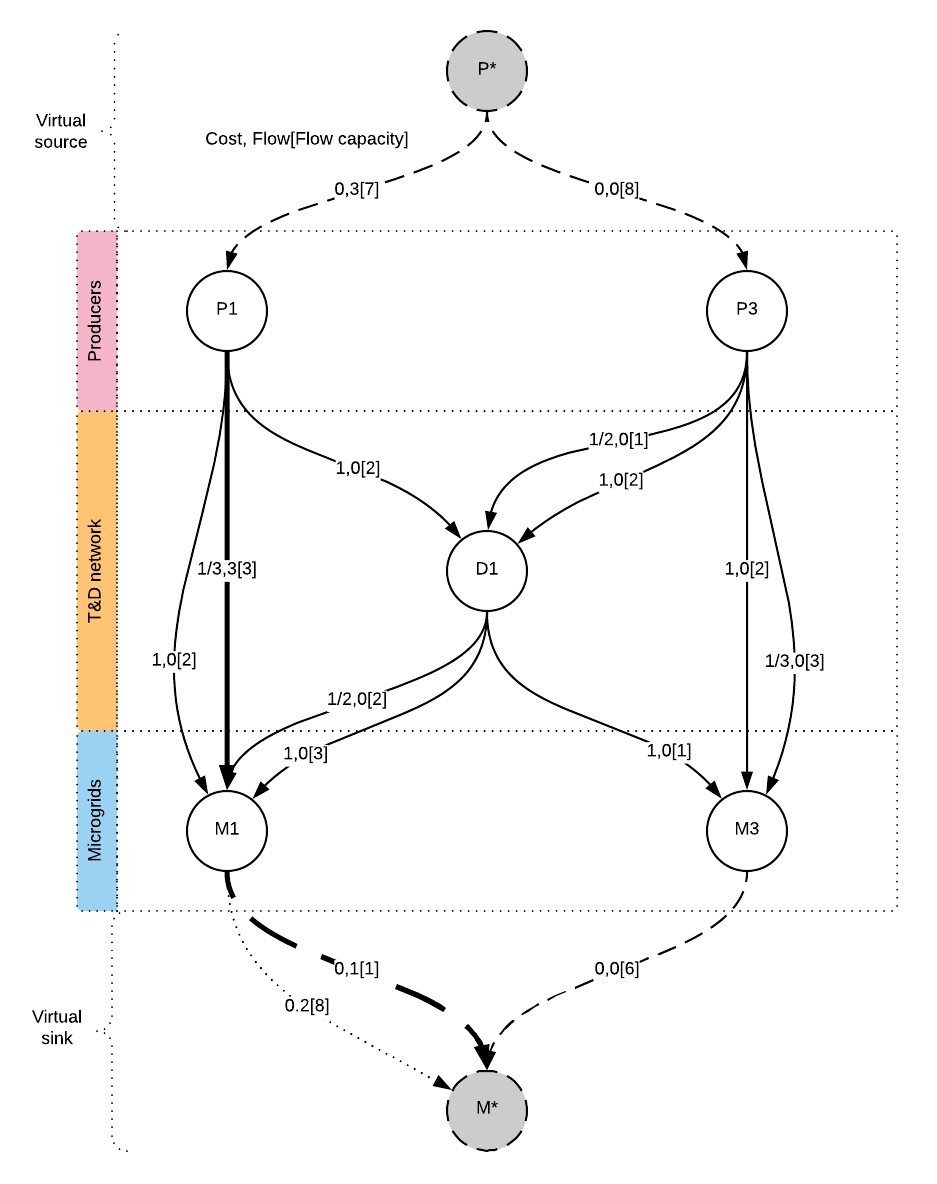}
\caption{Updating of routing.}
\label{rourou2}
\end{figure}

\begin{figure}[!ht]
\centering
\includegraphics [width=0.33\textwidth]{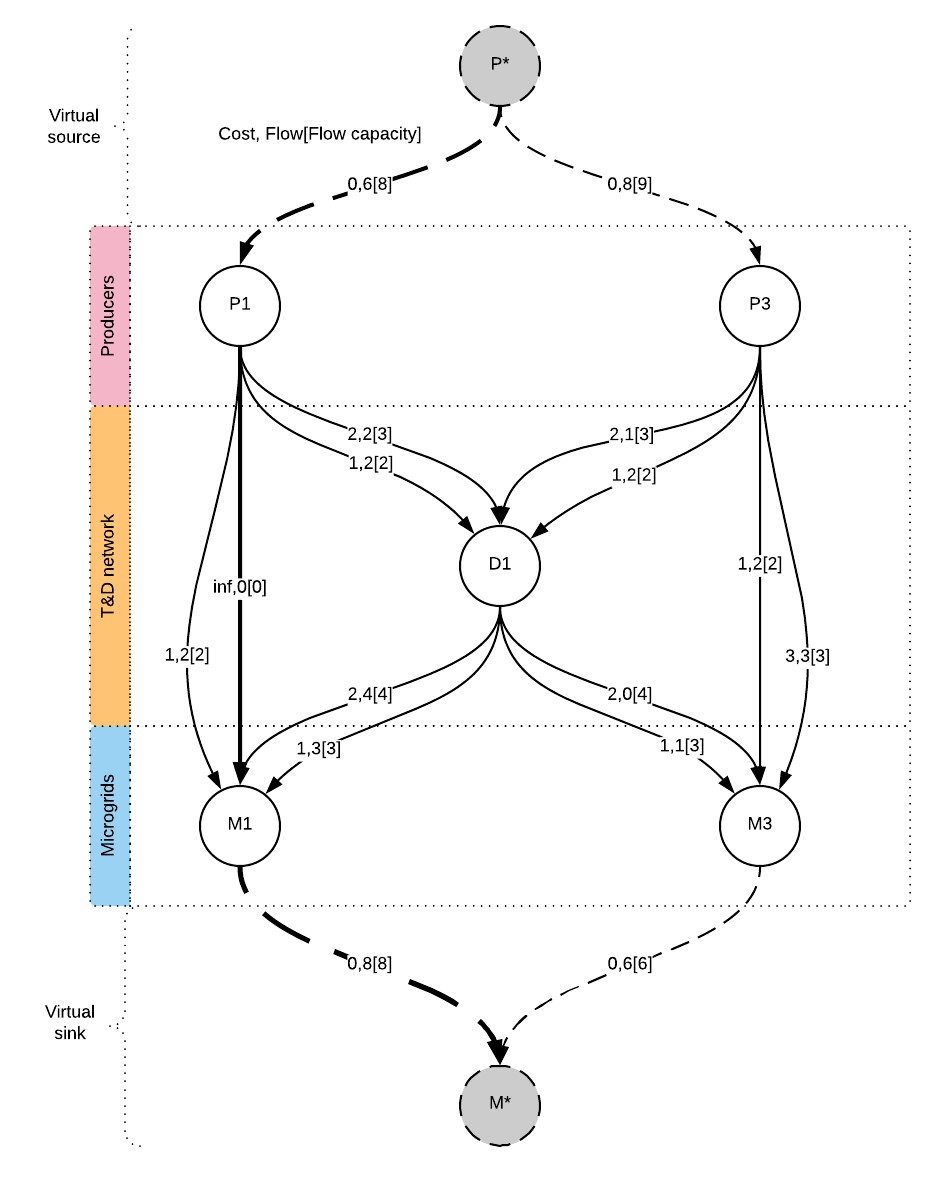}
\caption{New routing.}
\label{rourou3}
\end{figure}

At this point, the Sequence C starts. To avoid a fuzzy graph, the demands have been moderated. The graph from the figure \ref{rourou1} presents the routing at the previous iteration. At the current iteration, the line from the producer $P1$ to the microgrid $M1$ can no longer carry a standard load. Thus the line in bold in the figure \ref{rourou2} have a capacity of $0$. Once the highest cost route passing through this line is determined, the flow is removed from the first graph. The final graph is shown in the figure \ref{rourou3}. Since there is no more route from the source to the sink, an optimal solution is obtained.

All the energy is well routed, there is no need for feedbacks.

\subsection{A microgrid}
In this second simulation, the goal is focused on the homes interactions. In those tests, consumption's goal evolves in function of the hours of a day.

Simulations characteristics: 
\begin{itemize}
\item 5 homes
\item 15-25 devices per home
\item goal consumption:
\begin{itemize}
\item in $[\![10h,18h]\!]$, quadratic function whose the max is 5000
\item else, 3000 
\end{itemize} 
\item 5 simulated days
\end{itemize}

\begin{figure}[!ht]
\centering
\includegraphics[width=0.45\textwidth]{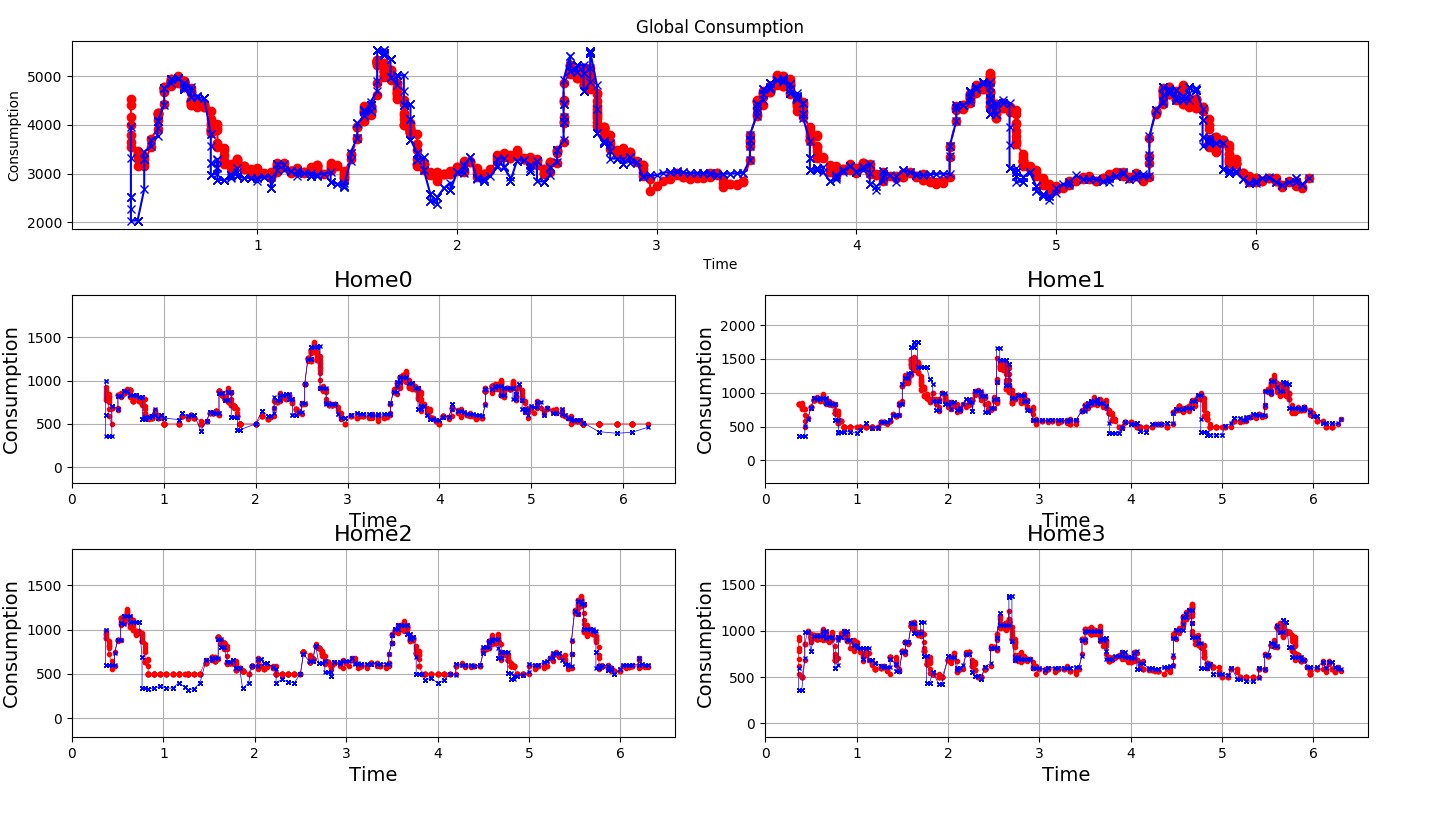}
\caption{Power production available in the neighborhood.}
\label{fig_consumption_test2}
\end{figure}

The regulation is affected according to their current consumption, to prevent too much consumption's losses. If the distribution is equitable, it is likely that it generates wastes (the consumptions are different from one house to another, planning devices varies from one house to another). We observe in the figure \ref{fig_consumption_test2} that the two curves are similar (the blue one represents the goal and the red one represents the consumption), which means that the neighborhood and home's consumption can be regulated within a goal.

Comparing the two results, it shows that the regulation can be regulated, according to a goal and users preferences. But it is meaningful to evaluate and catch errors and limitations. The following simulation explores its limits.

\subsection{Limits and future works}
\begin{figure}[!ht]
\centering
\includegraphics[width=0.40\textwidth]{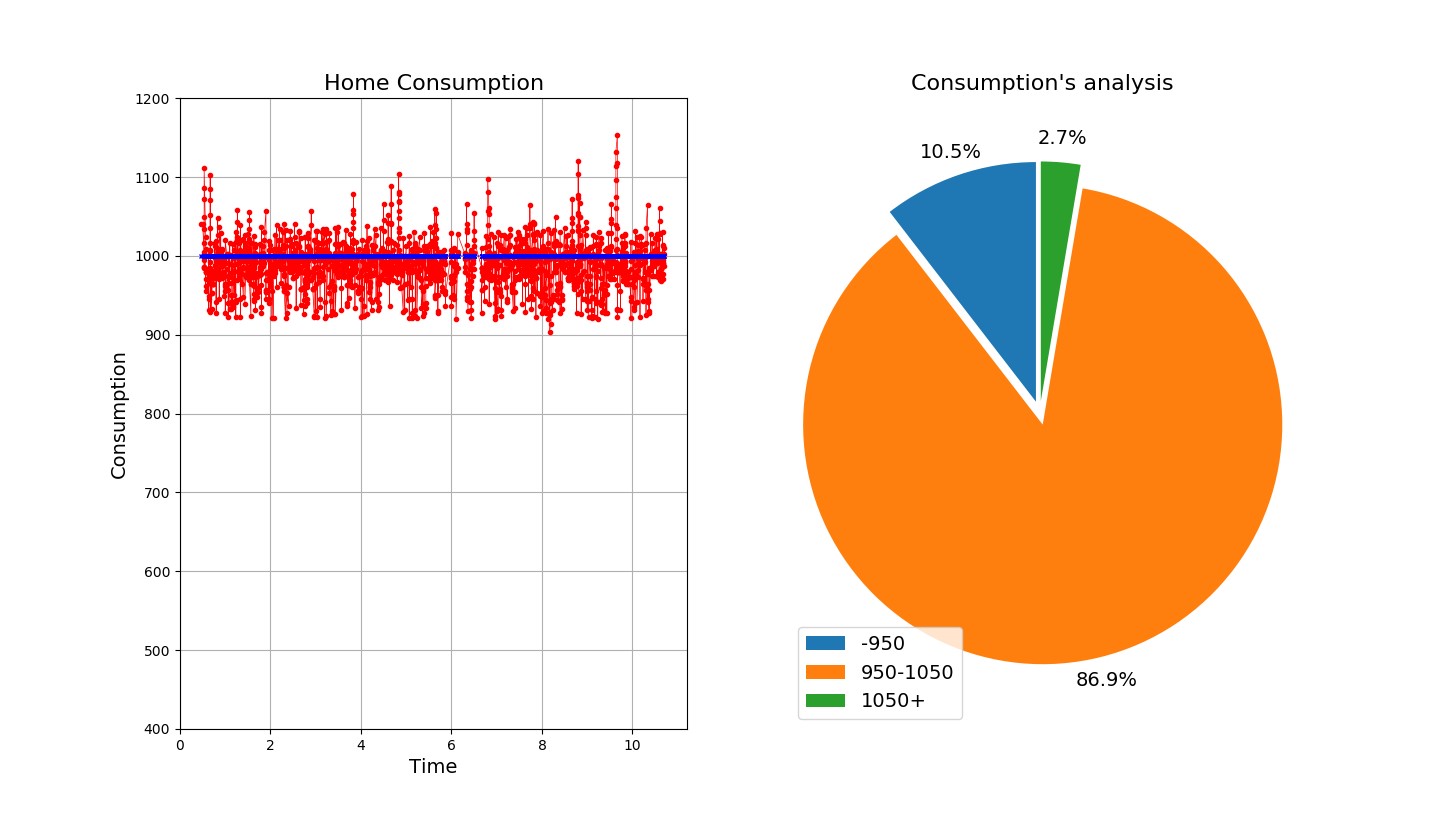}
\caption{Consumption for a 1 home simulation.}
\label{fig_consumption_home_test_one_home}
\end{figure}

The figure \ref{fig_consumption_home_test_one_home} presents the result of an isolated smart house, it is composed of two curves. The constant line represents the consumption's goal. The other one represents the home's consumption.

The average consumption on all test is 988, representing a 1.2\% error to the goal (1000). The minimum is 904 representing less of 10\% error and the maximum is 1153 representing 15\% error. 1696 values over 1953 are situated between 950 and 1050 representing more of 86\% of the values.

Those errors are explained as a result of a bad knowledge of the consumption behaviors. Indeed, the forecast does not take into account some pattern of consumption that occurs in a smart house or many of them.

In future works, a method to determine the most common patterns for each device consumption is experimented. This method is based on grammatical inference from the sequence of consumption of a device. This method builds a determined probabilistic finite automaton where the strategies of consumption are generated. From a current curve of consumption, the automaton is browsed, then a random walk builds various predictions. This method will be presented in a further paper.

\section{Conclusion}

In this paper, we addressed the optimization problem in power grids, especially in smart grids. The proposed model performs local optimization and feedback loops to reach a balanced optimum. The proposed smart grid's model provides some tools to simulate any smart grid. Grid network and energy sources change over time, in these conditions a model must be generic or it will be inadequate to adapt to the future. We note our model is straightforward and do not reflect the sheer complexity of a grid. However, new algorithms can be easily grafted to the model to enhance its flexibility, to represent a new kind of technology or to include new features.

%
\bibliographystyle{ACM-Reference-Format}
\bibliography{sample-base}

\end{document}